\documentclass{article}%
\usepackage{amsmath}
\usepackage{graphicx}%
\usepackage{amsfonts}%
\usepackage{amssymb}

\begin{document}

\title{Amplitude and phase of time \\
dependent Hamiltonian systems 
\\under the minimum uncertainty condition}
\date{}
\author{G. Landolfi\thanks{ e-mail: Giulio.Landolfi@le.infn.it},
 G. Ruggeri\thanks{
e-mail: Giovanna.Ruggeri@le.infn.it} \ and 
G. Soliani\thanks{ e-mail: Giulio.Soliani@le.infn.it}
\\\emph{Dipartimento di Fisica dell'Universit\`{a} di Lecce}
\\\emph{and INFN, Sezione di Lecce, I-73100 Lecce, Italy} }
\maketitle

\begin{abstract}
We investigate dynamical systems with time-dependent mass and frequency, with
particular attention on models attaining the minimum value of uncertainty
formula. A criterium of minimum uncertainty is presented and illustrated by
means of explicit and exactly solved examples.
The role of the Bogolubov coefficients, in
general and in the context of minimum uncertainty case, is discussed.

\hfill

PACS: 03.65.-w; 03.65.Ca; 42.50.Dv 

\end{abstract}

\section{Introduction}

\setcounter{equation}{0}

We analyze some peculiar situations emerging in the study of oscillators
characterized by constant and time-dependent mass and frequencies. Our
investigation is focused mainly on the treatment of a nonlinear auxiliary
equation which can be associated with the systems under consideration
([\ref{ermakov}-\ref{moya}]). A special attention is paid on the problem of
the formulation of the uncertainty relation, where in our framework the
auxiliary equation has a basic role.

A classical generalized or time-dependent oscillator (TDO) is governed by the
Hamiltonian%
\begin{equation}
H(t)=\frac{p^{2}}{2m}+\frac{1}{2}m\,\omega^{2}q^{2}\,\,,\label{CH1}%
\end{equation}
where $q,p$ are conjugate variables, and $m=m(t)$, $\omega=\omega(t)$ are
given functions of time. The Hamiltonian (\ref{CH1}) gives rise to the
equation of motion
\begin{equation}
\ddot{q}+M\,\dot{q}+\omega^{2}q=0\,\,,\label{eqtdo}%
\end{equation}
where $M=M(t)=\dot{m}\,/\,m$ and dot means time derivative. Through the
transformation
\begin{equation}
q=e^{-\frac{1}{2}\int_{t_{0}}^{t}M(t^{\prime})\,dt^{\prime}}%
\,y\,\,,\label{qqtilde}%
\end{equation}
Equation (\ref{eqtdo}) can be cast into the form%
\begin{equation}
\ddot{y}+\Omega^{2}(t)\,y=0\,\,,\label{eqqtilde}%
\end{equation}
where%
\begin{equation}
\Omega^{2}(t)=\frac{1}{4}\left(  4\omega^{2}-2\dot{M}-M^{2}\right)
\,\,.\label{defOmega2}%
\end{equation}
The following equation%
\begin{equation}
\ddot{\sigma}+\Omega^{2}(t)\,\sigma=\frac{K}{\sigma^{3}}%
\,\,,\label{eqausiliaria}%
\end{equation}
where $K$ is a constant, can be related to Eq. (\ref{eqqtilde}), in the
sense that if $y_{1}$ and $y_{2}$ are two independent solutions of Eq.
(\ref{eqqtilde}), then the general solution of the auxiliary equation
(\ref{eqausiliaria}) can be written as [\ref{eliezer}]%

\begin{equation}
\sigma=(Ay_{1}^{2}+By_{2}^{2}+2Cy_{1}y_{2})^{\frac{1}{2}}%
\,\,,\label{sigmasqrt}%
\end{equation}
$A,B,C$ being constants such that%

\begin{equation}
AB-C^{2}=\frac{K}{W_{0}^{2}} \label{eqABCW0}%
\end{equation}
and $W_{0}=y_{1}\dot{y}_{2}-\dot{y}_{1}y_{2}=const$ is the Wronskian. From the
theory of the auxiliary equation (\ref{eqausiliaria}) a phase is involved
given by a real function $\theta(t)$ defined by%

\begin{equation}
\theta(t)=\int_{t_{0}}^{t}\frac{dt^{\prime}}{\sigma^{2}(t^{\prime}%
)}\,\,\,.\label{fasetheta}%
\end{equation}
(See [\ref{ermakov}-\ref{pinney},\ref{goff}]; for some applications:
[\ref{lewis}], [\ref{profilo}]).

In this Letter we discuss some aspects of the quantum theory of generalized
oscillators with time-dependent mass and frequency. Precisely, we revise the
formulation of the uncertainty relation in terms of solutions of the auxiliary
equation (\ref{eqausiliaria}), with a particular care to the analysis of the
Bogolubov coefficients and the possibility to attain the minimum value of the
uncertainty product $\left(  \Delta_{\alpha}Q\right)  \,\,\left(
\Delta_{\alpha}P\right)  $, where $\Delta_{\alpha}Q\,\,$\ and $\Delta_{\alpha
}P$ denote the variances of the position and momentum operators between
extended coherent states. 
A simple criterium is found which allows one to build up time-dependent
models reaching the minimum uncertainty relation. Some examples of these
systems are presented. 
One of these, which leads to an equation of the Bessel-type, 
is associated with an auxiliary equation (see (\ref{Y4})) exactly solved
in terms of a (convergent) power series expansion.

The outline is as follows. In Section 2 some examples of (real)
exact solutions of the auxiliary equation (\ref{eqausiliaria}) are shown. In
Section 3, the uncertainty relation in terms of the Bogolubov coefficients
of the generalized oscillator with time-dependent mass and frequency is
considered. A criterium of minimum uncertainty relation is also presented.
Section 4 concerns with examples of exactly solvable TDO's minimizing the
uncertainty formula. In Section 5 a concluding discussion is reported.

\section{Examples of exact solutions of Eq. (\ref{eqausiliaria})}

\setcounter{equation}{0}

Two interesting cases where Eq. (\ref{eqausiliaria}) can be exactly solved via
(\ref{sigmasqrt}), are represented by a) the usual (time-independent) harmonic
oscillator and b) the Kanai-Caldirola oscillator. In what follows, we shall
deal with both cases.

\begin{description}
\item[Case a)] \underline{The harmonic oscillator}
\end{description}

Let us assume that $\omega(t)$ and $m(t)$ are constants, namely $\omega
(t)=\omega_{0}$ and $m(t)=m_{0}$. Then, the Hamiltonian (\ref{CH1}) takes the
form
\begin{equation}
H(t)=\frac{p^{2}}{2m_{0}}+\frac{1}{2}m_{0}\,\omega_{0}^{2}\,q^{2}\,\,\,,
\end{equation}
while the auxiliary equation (\ref{eliezer}) becomes%
\begin{equation}
\ddot{\sigma}+\omega_{0}^{2}\,\sigma=\frac{K}{\sigma^{3}}\,\,.
\end{equation}
The general solution of this equation can be obtained via (\ref{sigmasqrt}) by
choosing, say, $q_{1}=q_0 \cos\omega_{0}t$, 
$q_{2}=q_0 \sin\omega_{0}t$ . If we put
$K=\frac{1}{4}$ for convenience, and hereafter we limit ourselves to the
coefficients $A,B,C$\thinspace$\ $leading to real solution $\sigma(t)$, then
equation (\ref{eqABCW0}) is satisfied by%
\begin{equation}
A=B=\frac{1}{2\omega_{0}},\quad C=0\label{ABCset1}%
\end{equation}
for $k=\omega_{0}$ ($k$ is defined below), and%

\begin{align}
A &  =\frac{k}{2\omega_{0}^{2} q_0^2}-
\frac{\sqrt{k^{2}-\omega_{0}^{2}}}{2\omega
_{0}^{2} q_0^2}\cos(2c_{1}\pm2\omega_{0}t_0)\,\,\,,\label{Aset2}\\
B &  =\frac{k}{2\omega_{0}^{2} q_0^2}+
\frac{\sqrt{k^{2}-\omega_{0}^{2}}}{2\omega_{0}^{2} q_0^2}
\cos(2c_{1}\pm2\omega_{0}t_0)\,\,\,,\label{Bset2}\\
C &  =\mp\frac{\sqrt{k^{2}-\omega_{0}^{2}}}{2\omega_{0}^{2}q_0^2}
\sin(2c_{1}\pm 2\omega_{0}t_0)\,\,\,,\label{Cset2}%
\end{align}
where $c_{1}$ is an arbitrary constant, for $k>\omega_{0}$, where $k$ is a
constant of integration appearing in the equation%
\begin{equation}
\dot{\sigma}^{2}+\omega_{0}^{2}
\sigma^2 +\frac{1}{4\sigma^{2}}=k\qquad\label{eq27}%
\end{equation}
arising from (\ref{eqausiliaria}).
Inserting the coefficients (\ref{ABCset1}) and (\ref{Aset2})-(\ref{Cset2})
into (\ref{sigmasqrt}) provides%
\begin{equation}
\sigma=\frac{1}{\sqrt{2\omega_{0}}}\,\,\,,\label{sigmaset1}%
\end{equation}
and
\begin{equation}
\sigma=\left[  \frac{k}{2\omega_{0}^{2}}-\frac{\sqrt{k^{2}-\omega_{0}^{2}}%
}{2\omega_{0}^{2}}\cos(2c_{1}\pm2\omega_{0}t)\right]  ^{\frac{1}{2}%
}\,\,\,,\label{sigmaset2}%
\end{equation}
respectively. The phases (\ref{fasetheta}) corresponding to the solutions
(\ref{sigmaset1}) and (\ref{sigmaset2}) are given by%
\begin{equation}
\theta(t)=2\omega_{0}\left(  t-t_{0}\right)
\end{equation}
and%
\begin{equation}
\theta(t)=\left.  \pm2\tan^{-1}\left[  \frac{1}{\omega_{0}}\left(
k+\sqrt{k^{2}-\omega_{0}^{2}}\right)  \,\tan\left(  c_{1}\pm\omega
_{0}t^{\prime}\right)  \right]  \right|  _{t_{0}}^{t}%
\end{equation}
respectively.

\begin{description}
\item[Case b)] \underline{The Kanai-Caldirola oscillator}
\end{description}

Now let us suppose that the frequency $\Omega^{2}(t)$ defined in
(\ref{defOmega2}) is a constant, \textit{i.e.} $\Omega^{2}(t)=\Omega_{0}^{2}$,
and $m(t)=m_{0}\exp(\gamma t)$ , where $\gamma\equiv M_{0}$ is the damping. So
the equation of motion (\ref{eqtdo}) corresponds to a particular form of the
Kanai-Caldirola Hamiltonian [\ref{kanai}-\ref{caldirola}], \textit{i.e.}%
\begin{equation}
H=\frac{p^{2}}{2m_{0}}\exp\left(  -\gamma\,t\right)  +\frac{1}{2}m_{0}%
\exp\left(  \gamma\,t\right)  \,\,\Omega_{0}^{2}\,q^{2}%
\,\,\,.\label{kanaicaldirola}%
\end{equation}
The auxiliary equation (\ref{eqausiliaria}) can be written as%
\begin{equation}
\ddot{\sigma}+\Omega_{0}^{2}\sigma=\frac{1}{4\sigma^{3}}\,\,.
\end{equation}
Since from (\ref{defOmega2}) we get
\begin{equation}
\Omega^{2}(t)=\Omega_{0}^{2}=\omega_{0}^{2}-\frac{\gamma^{2}}{4}\,\,,
\end{equation}
we have two possible situations: \textit{i) }$\Omega_{0}^{2}>0$ and
\textit{ii)} $\Omega_{0}^{2}<0$ . In case \textit{i)} two independent
solutions of the equation of motion (\ref{eqtdo}) are $q_{1}=q_{0}\cos
\Omega_{0}t$ , $q_{2}=q_{0}\sin\Omega_{0}t$ . Consequently, case \textit{i)}
is similar to the case $M=0$ considered previously. Therefore, for $\sigma$ we
find two possible solutions, which coincide with (\ref{sigmaset1}) and
(\ref{sigmaset2}) where $\omega_{0}$ is replaced by $\Omega_{0}$. Conversely,
case \textit{ii)} is concerned with the solution
\begin{equation}
\sigma=\left[  c_{1}+\sqrt{c_{1}^{2}+\frac{1}{4\left|  \Omega_{0}\right|
^{2}}}\,\,\,\cosh\left(  2c_{2}\pm2\left|  \Omega_{0}\right|  \,t\right)
\right]  ^{\frac{1}{2}}\label{sigmaset3}%
\end{equation}
which implies the phase
\begin{equation}
\theta(t)=\left.  \pm2\tan^{-1}\left[  \left(  \sqrt{4\left|  \Omega
_{0}\right|  ^{2}c_{1}^{2}+1}-2\left|  \Omega_{0}\right|  c_{1}\right)
\,\tanh\left(  c_{2}\pm\left|  \Omega_{0}\right|  \,t^{\prime}\right)
\right]  \right|  _{t_{0}}^{t}\,\,\,.
\end{equation}

\section{The uncertainty relation in terms of the Bogolubov coefficients for
the TDO}

\setcounter{equation}{0}

In [\ref{profilo}], in the context of the determination of Noether invariant
operators and dynamical group of the TDO described by the Hamiltonian
operator%
\begin{equation}
\hat{H}=\frac{P^{2}}{2m}+\frac{1}{2}m\,\omega^{2}(t)\,Q^{2}\,\,\,,\label{HQP}%
\end{equation}
the following time-dependent lowering and raising operators $a(t)$ and
$a^{\dag}(t)$ have been introduced:%

\begin{align}
a(t) &  =\left(  \frac{m}{\hbar}\right)  ^{\frac{1}{2}}\left\{  \frac
{Q}{2\sigma}+i\left[  \frac{\sigma}{m}P-\left(  \dot{\sigma}-\frac{M}{2}%
\sigma\right)  Q\right]  \right\}  \;\,,\label{aQP}\\
a^{\dag}(t) &  =\left(  \frac{m}{\hbar}\right)  ^{\frac{1}{2}}\left\{
\frac{Q}{2\sigma}-i\left[  \frac{\sigma}{m}P-\left(  \dot{\sigma}-\frac{M}%
{2}\sigma\right)  Q\right]  \right\}  \,\,\,,\label{a+QP}%
\end{align}
where $m=m(t)$, $M=\frac{\dot{m}}{m}$, $[Q,P]=i\hbar$, $[a(t),a^{\dag
}(t)]=\hat{1}$, and%

\begin{equation}
\ddot{\sigma}+\Omega^{2}\sigma=\frac{1}{4\sigma^{3}}\,\,\,,\label{sigma14}%
\end{equation}
where $\Omega^{2}$ is given by (\ref{defOmega2}). Now, from (\ref{sigma14}) we
obtain%
\begin{equation}
\dot{\sigma}^{2}+\Omega^{2}\sigma^{2}+\frac{1}{4\sigma^{2}}=k+F\,\,\,,
\end{equation}
where $F$ is defined by%
\begin{equation}
F=\int^{t}\Omega\,\dot{\Omega}\,\sigma^{2}\,dt^{\prime}=\frac{1}{4}\int
^{t}\sigma^{2}\left(  4\,\omega\,\dot{\omega}-\ddot{M}-M\,\dot{M}\right)
\,dt^{\prime}\,\ .
\end{equation}
Form (\ref{aQP}) and (\ref{a+QP}) we find
\begin{equation}
Q=\sqrt{\frac{\hbar}{m}}\sigma(a+a^{\dagger})\qquad,\qquad P=\sqrt{\hbar
m}\left(  \xi a+\xi^{\ast}a^{\dagger}\right)  \label{QP}%
\end{equation}
where%
\begin{equation}
\xi=-\left[  \frac{i}{2\sigma}+\left(  \frac{M}{2}\sigma-\dot{\sigma}\right)
\right]  \,\,\,,\label{xi}%
\end{equation}
and%
\begin{equation}
\left|  \xi\right|  ^{2}=\left[  \frac{1}{4}\sigma^{2}+\left(  \frac{M}%
{2}\sigma-\dot{\sigma}\right)  ^{2}\right]  =\left[  k+F+\left(  \frac{M^{2}%
}{4}-\Omega^{2}\right)  \sigma^{2}-M\sigma\dot{\sigma}\right]  \;.\label{xi2}%
\end{equation}
Now, let us deal with the variances%
\begin{equation}
\left(  \Delta_{\alpha}Q\right)  ^{2}=\left\langle \,Q^{2}\,\right\rangle
-\left\langle \,Q\,\right\rangle ^{2}\,\quad,\quad\left(  \Delta_{\alpha
}P\right)  ^{2}=\left\langle \,P^{2}\,\right\rangle -\left\langle
\,P\,\right\rangle ^{2}%
\end{equation}
where the expectation value is given by
$\left\langle \,\ldots\right\rangle =\left\langle
\alpha\left|  \ldots\right|  \alpha\,\right\rangle $ , $\left|  \alpha
\right\rangle $ denoting the generalized coherent state. Thus, we have
\begin{equation}
\left(  \Delta_{\alpha}Q\right)  ^{2}=\frac{\hbar}{m}\sigma^{2}\,\,,\quad
\left(  \Delta_{\alpha}P\right)  ^{2}=\hbar\,m\,\,\left|  \xi\right|
^{2}\,\,,
\end{equation}
which entail the uncertainty formula%
\begin{align}
\left( \Delta_{\alpha}Q \right)
\,\,\left(\Delta_{\alpha}P \right)&  =\frac{\hbar}{2}\left[  1+4\sigma
^{2}\left(  \dot{\sigma}-\frac{M}{2}\sigma\right)  ^{2}\right]  ^{\frac{1}{2}%
}=\nonumber\\
&  =\hbar\sigma\left[  k+F+\left(  \frac{M^{2}}{4}-\Omega^{2}\right)
\sigma^{2}-M\sigma\dot{\sigma}\right]  ^{\frac{1}{2}}\quad
\label{indeterminazioneQP}%
\end{align}
which holds for a generalized oscillator with time-dependent mass and frequency.

For $m=m_{0}$ (\textit{i.e.} $M=0$), Eq. (\ref{indeterminazioneQP}) reproduces
the relation (see (\ref{xi2}))%
\begin{equation}
\left(\Delta_{\alpha}Q
\right)\,\,\left(\Delta_{\alpha}P\right)
=\frac{\hbar}{2}\sqrt{1+4\sigma^{2}%
\dot{\sigma}^{2}}\geq\frac{\hbar}{2}\quad,
\end{equation}
where the minimum is attained only for $\dot{\sigma}=0$.

In order to write the Bogolubov transformation, we introduce the operators%
\begin{equation}
Q=\sqrt{\frac{\hbar}{2m_{0}\omega_{0}}}(a_{0}+a_{0}^{\dagger})\qquad,\qquad
P=-i\sqrt{\frac{\hbar m_{0}\omega_{0}}{2}}\left(  a_{0}-a_{0}^{\dagger
}\right)  \label{QP0}%
\end{equation}
where $a_{0}=a_{0}(0)$ is the Schr\"{o}dinger picture fixed photon annihilator
operator. Substitution from (\ref{QP0}) into (\ref{aQP}) and (\ref{a+QP})
provides%
\begin{equation}
a(t)=\mu(t)\,a_{0}+\nu(t)\,a_{0}^{\dagger}\quad,\label{amunua0}%
\end{equation}
where the Bogolubov coefficients $\mu(t)$ and $\nu(t)$ take the form%
\begin{equation}
\mu(t)=\sqrt{\frac{m}{2m_{0}\omega_{0}}}\left[  \eta+\frac{m_{0}\omega_{0}}%
{m}\sigma\right]  \quad,\quad\nu(t)=\sqrt{\frac{m}{2m_{0}\omega_{0}}}\left[
\eta-\frac{m_{0}\omega_{0}}{m}\sigma\right]  \label{munu}%
\end{equation}
with
\begin{equation}
\eta=-i\xi^{\ast}=\frac{1}{2\sigma}+i\left(  \frac{M}{2}\sigma-\dot{\sigma
}\right)  \,\,\ .
\end{equation}
For any real solution of Eq. (\ref{sigma14}) we get
\begin{align}
\left|  \mu(t)\right|  ^{2} &  
=\frac{m}{2m_{0}\omega_{0}}\left\{  k+F+\left(
\frac{m_{0}^{2}\omega_{0}^{2}}{m^{2}}+\frac{M^{2}}{2}-\omega^{2}-\frac{\dot
{M}}{2}\right)  \,\sigma^{2}-M\sigma\dot{\sigma}\right\}  +\frac{1}%
{2}\label{mu2}\\
\left|  \nu(t)\right|  ^{2} &  
=\frac{m}{2m_{0}\omega_{0}}\left\{  k+F+\left(
\frac{m_{0}^{2}\omega_{0}^{2}}{m^{2}}+\frac{M^{2}}{2}-\omega^{2}-\frac{\dot
{M}}{2}\right)  \,\sigma^{2}-M\sigma\dot{\sigma}\right\}  -\frac{1}%
{2}\label{nu2}%
\end{align}
so that the complex functions $\mu(t)$ and $\nu(t)$ possess the property%
\begin{equation}
\left|  \mu(t)\right|  ^{2}-\left|  \nu(t)\right|  ^{2}=1\,\,\,.
\end{equation}
Furthermore, the quantities%
\begin{equation}
\mu(t)+\nu(t)=
\sqrt{\frac{2m}{m_{0}\omega_{0}}}\eta\qquad,\qquad
\mu(t)-\nu(t)=
\sqrt{\frac{2m_{0}\omega_{0}}{m}}\sigma
\end{equation}
can be involved in the uncertainty formula (\ref{indeterminazioneQP}) ,
\textit{i.e.}%
\begin{align}
\left( \Delta_{\alpha}Q \right)\,
\left( \Delta_{\alpha}P \right)
&  =\frac{\hbar}{2}\left[  1+
4\sigma^{2}\left(  \dot{\sigma}-
\frac{M}{2}\sigma\right)  ^{2}\right]  ^{\frac{1}{2}}=\nonumber\\
&  =\frac{\hbar}{2}\,\left(  4\,\sigma^{2}\left|  \eta\right|^{2}\right)
^{\frac{1}{2}}
=\frac{\hbar}{2}\left|  \mu(t)+\nu(t)
\right|  \left|  \mu(t)-\nu(t)\right| 
\geq\frac{\hbar}{2}\,\,\,.\label{incertezzabogo}%
\end{align}
The uncertainty formula (\ref{incertezzabogo}) is closely related to the
concept of \textit{coherent states} for the generalized oscillators. We remind
the reader that such coherent states in the context of Lewis-Riesenfeld theory
were constructed by Hartley and Ray in 1982 [\ref{lewis}]. These states share
all the features of the coherent states of the conventional (time-independent)
oscillator except that the uncertainty formula, \textit{i.e.} the product of
position and momentum is \textit{not} minimum. A few years later, Pedrosa
[\ref{pedrosa}] showed that the coherent states devised by Hartley and Ray for
the TDO are actually equivalent to the well-known \textit{squeezed} states
(see, for instance, [\ref{yuen}] and [\ref{walls}]).

A criterium of minimum uncertainty for a time-dependent oscillator with
variable mass and frequency is expounded in next Subsection.

\subsection{A criterium of minimum uncertainty relation for the TDO}

The product (\ref{indeterminazioneQP}) reaches its minimum value whenever the
condition
\begin{equation}
\frac{M}{2}\sigma=\dot{\sigma}%
\end{equation}
is fulfilled. This implies the constraint
\begin{equation}
\sigma=cm^{1/\,2}\,\,\,,\label{sigmamin}%
\end{equation}
where $c$ is a constant of integration. By introducing (\ref{sigmamin}) 
in the auxiliary equation (\ref{eqausiliaria}) yields%
\begin{equation}
2\,m\,\ddot{m}-\dot{m}^{2}+4\,\Omega^{2}m^{2}=\frac{1}{c^{4}}%
\,\,\,.\label{mmin}%
\end{equation}
This equation can be elaborated as follows. Substitution from
\begin{equation}
m\,\ddot{m}=m^{2}\frac{d}{dt}\frac{\dot{m}}{m}+\dot{m}^{2}%
\end{equation}
into (\ref{mmin}), we find%
\begin{equation}
m(t)\,\omega(t)=\frac{1}{2c^{2}}\label{momegac2}%
\end{equation}
by virtue of (\ref{defOmega2}). Equation (\ref{momegac2}) represents a
\textit{criterium} of minimum uncertainty for the product of variances
$(\Delta_{\alpha}Q)\,\,(\Delta_{\alpha}P)$ for the generalized oscillator
(\ref{eqtdo}). That is, 
an oscillator with generally time-dependent mass and frequency
minimizing the uncertainty formula is described by the equation%
\begin{equation}
\ddot{q}-\frac{\dot{\omega}}{\omega}\dot{q}+\omega^{2}q=0\,\,\,.
\end{equation}
In the light of this result, it is clear once again that for the usual
harmonic oscillator (corresponding to the particular case in which
$\omega(t)=\omega_{0}=const$ and $m=m_{0}=(2\omega_{0}c^{2})^{-1}$), 
the unique exact solution of the auxiliary equation
minimizing the uncertainty product is $\sigma=\frac{1}{\sqrt{2\omega_{0}}}$ .
The other solution (\ref{sigmaset2}) can minimize the uncertainty formula only
approximately, say for $\omega_{0}^{2}<k<<3\omega_{0}^{2}$.

Notice that in the minimum uncertainty case the Eq. (\ref{fasetheta}) 
takes the form
\begin{equation}
\theta(t)=2 \int_{t_{0}}^{t} \omega(t^{\prime}) {dt^{\prime}}\,\,\,.
\label{fasethetaminimum}
\end{equation}

\section{Examples of exactly solvable TDO's minimizing the uncertainty formula}

\setcounter{equation}{0}

Below we shall display a few interesting examples of TDO's which correspond to
the minimum of the uncertainty formula.

\begin{description}
\item[Case I)] 
\underline{A TDO with exponentially decreasing frequency}
\end{description}

Let us consider a TDO with frequency
\begin{equation}
\omega=\omega_{0}\exp\left(  -\gamma_{0}t\right)  \,\,\,,
\end{equation}
and mass
\begin{equation}
m=\frac{1}{2c^{2}\omega_{0}}\,\exp(\gamma_{0}t)\,\,\,,
\end{equation}
so that the criterium (\ref{momegac2}) is verified ($\gamma_{0}=const>0$).
The equation of motion (\ref{eqtdo}) reads%
\begin{equation}
\ddot{q}+\gamma_0 \dot{q}+\omega_{0}^{2}\exp\left(  -2\gamma_{0}\,t\right)
\,\,q=0\,\,\,.\label{x37}%
\end{equation}
By means of the transformation (\ref{qqtilde}), \textit{i.e.}%
\thinspace $q=\exp\left(  \frac{\gamma_{0}}{2}t\right) \,y$, 
Equation (\ref{x37}) can be cast into the equation
\begin{equation}
\ddot{y}+\Omega^{2}(t)\,y=0 \quad ,\quad
\Omega^{2}(t)=\omega_{0}^{2}\,\exp\left(  -2\gamma_{0}\,t\right)
-\frac{\gamma_{0}^{2}}{4} \quad,
\label{yddott}
\end{equation}
as one finds from (\ref{defOmega2}). The general solution of 
Equation (\ref{yddott}) can be given in terms of elementary functions:
\[
y=\sqrt{\gamma_0} e^{\gamma_{0}\,t}
\left[ c_{1}\,\cos \left( \frac{\omega_{0}}{\gamma_0} e^{-\gamma_{0}\,t}
\right) +
c_{2}\sin \left( \frac{\omega_{0}}{\gamma_0} e^{-\gamma_{0}\,t} \right) 
\right] \,\,,
\]
where $c_{1}$ and $c_{2}$ are arbitrary constants. 

Finally, due to Eq. (\ref{fasethetaminimum}), the phase is 
$\theta=\left. -2 \frac{\omega_{0}}{\gamma_0} e^{  -\gamma_{0}t}  \,
\right|^t_{t_0} $.

\begin{description}
\item[Case II)] 
\underline{A TDO with mass $m=m_{0}t^{2}$ and frequency 
$\omega=\frac{1}{2m_{0}c^{2}t^{2}}$}
\end{description}

In this case the equation of motion (\ref{eqtdo})\bigskip\ is%
\begin{equation}
\ddot{q}+\frac{2}{t}\dot{q}+\frac{1}{(2m_{0}c^{2}t^{2})^{2}}\,q=0\,\,\,,
\label{xx47}
\end{equation}
which may be transformed into
\begin{equation}
\ddot{y}+\frac{1}{(2m_{0}c^{2}t^{2})^{2}}\,y=0\label{x47}%
\end{equation}
via $q=\frac{y}{t}$. Equation (\ref{xx47}) allows therefore 
the general solution
\begin{equation}
q=\left[  c_{1}\cos\frac{1}{(2m_{0}c^{2})\,t}+c_{2}\sin\frac{1}{(2m_{0}%
c^{2})\,t}\right]  \,\,\,\, ,
\end{equation}
$c_{1}$, $c_{2}$ being arbitrary constants. From Eq. (\ref{fasethetaminimum})
we also get $\theta=\left. \frac{-1}{m_{0}c^{2}t} \right| ^t_{t_0}$.  

\begin{description}
\item[Case III)] 
\underline{A TDO described by an equation of the Bessel-type}
\end{description}

Let us suppose that
\begin{equation}
m=m_{0}\,\alpha(t) \qquad , \qquad
\omega=\frac{\omega_{0}}{\alpha(t)}\,\,.
\label{malpha}%
\end{equation}
Then, $m\,\omega=m_{0}\,\omega_{0}=(2c^{2})^{-1}$. If we demand that
\begin{equation}
\Omega^{2}(t)=\Omega_{0}^{2}\,\left(  k_{0}^{2}+\frac{\nu^{2}}{t^{2}}\right)
\label{Y3}%
\end{equation}
where $k_{0}$,$\nu$ are constants, substitution from (\ref{malpha}) 
into (\ref{Y3}) yields the condition
\begin{equation}
2\alpha\ddot{\alpha}-\dot{\alpha}^{2}-4\omega_{0}^{2}+4\,\alpha^{2}%
\,\Omega_{0}^{2}\,\left(  k_{0}^{2}+\frac{\nu^{2}}{t^{2}}\right)
=0\,\,\,.\label{Y4}%
\end{equation}
This nonlinear equation can be exactly solved in terms of the series expansion
(see the Appendix)%
\begin{equation}
\alpha(t)=\sum_{k=0}^{\infty}a_{2k+1}\,t^{2k+1}\,\,\,,\label{alphatak}%
\end{equation}
where all the coefficients of even index, $a_{2k}$, are zero and
\begin{equation}
a_{2k+1}=(-1)^{k}\,\mu^{2k}\,\prod_{j}^{2k+1}\,\frac{[2(k+1)-j]}%
{(2k+3-j)\,[j(j-2)+\lambda^{2}]}\,\,a_{1}\,\,\,,\label{a2k+1a1}%
\end{equation}
where
\begin{equation}
a_{1}=\frac{2\omega_{0}}{\sqrt{\lambda^{2}-1}}\quad,\quad\lambda=2\,\Omega
_{0}\,\nu\quad,\quad\mu=2\,\Omega_{0}\,k_{0}\quad.\label{Y7}%
\end{equation}
Provided that $\lambda^{2}>1$, we thus get
\begin{equation}
m=m_{0}\sum_{k=0}^{\infty}a_{2k+1}\,t^{2k+1}\quad,\quad\omega=\frac{\omega
_{0}}{a_1 t}\sum_{k=0}^{\infty}\tilde{a}_{2k}\,t^{2k}%
\end{equation}
where (see \textit{e.g.} [\ref{gradshteyn}])
\begin{align*}
\tilde{a}_{k} 
&  =\left( \frac{-1}{a_1} \right)^{k} \left|
\begin{array}
[c]{cccccc}%
0 & a_{1} & 0 & 0 & 0 & 0\\
a_{3} & 0 & a_{1} & 0 & 0 & 0\\
0 & a_{3} & 0 & a_{1} & 0 & 0\\
a_{5} & 0 & a_{3} & 0 & \dots & \dots\\
\dots & \dots & \dots & \dots & \dots & a_{1}\\
a_{k+1} & a_{k} & a_{k-1} & \dots & \dots & 0
\end{array}
\right|  \,\,\,\,\,.
\end{align*}
It is worthwhile noting that for $k_{0}^{2}>>\nu^{2}/t^{2}$ one would get from
(\ref{Y4}) the solution
\[
\alpha=c_{1}+\sqrt{c_{1}^{2}-\frac{\omega_{0}^{2}}{\Omega_{0}^{2}\,k_{0}^{2}}%
}\,\sin\left[  \sqrt{4\Omega_{0}^{2}k_{0}^{2}}\,(t+c_{2})\right]
\]
where $c_{1}$ and $c_{2}$ are constants. Inserting such a solution into
(\ref{malpha}), (\ref{Y4}), and solving the equation of
motion (\ref{eqtdo}) we would obtain
\begin{equation}
q(t)=C_{1}\sin\left\{  \sqrt{
\frac{\Omega_{0}^{2}}{\omega_0^2}}
\tan^{-1}\left[ \left(
\sqrt{\frac{c_{1}^{2}\Omega_{0}^{2}\,k_{0}^{2}}{\omega_{0}^{2}}-1}
-c_1 \sqrt{\frac{\Omega_0^2 k_0^2}{\omega_0^2}} \right)
\tan\sqrt{\Omega_{0}^{2}k_{0}^{2}}%
\,(t+c_{2})\right]  +\phi_{0}\right\}
\end{equation}
where $C_{1}$ and $\phi_{0}$ are real constants. 
Due to (\ref{fasethetaminimum}), the phase $\theta$ now reads
\begin{equation}
\theta=2 \omega_0 \left. \left[  \frac{1}{{a}_1} \ln t
+ \sum_{k=1}^{\infty}\frac{\tilde{a}_{2k}}{2k}\,t^{2k}%
\right] \right|^t_{t_0}
\end{equation}

In the special case in which $\mu=0$ ($k_{0}=0$), we simply have%
\begin{equation}
\alpha(t)=a_{1}\,t\,\,\,.\label{a1t}%
\end{equation}
Inserting (\ref{a1t}) in (\ref{Y4}) provides just the condition (\ref{Y7}),
while the equation of motion (\ref{eqtdo}) becomes%
\begin{equation}
\ddot{q}+\frac{1}{t}\dot{q}+\left(  \Omega_{0}^{2}\,\nu^{2}-\frac{1}%
{4}\right)  \frac{1}{t^{2}}\,q=0\,\,\,,
\end{equation}
which affords the general solution
\begin{equation}
q=c_{1}\,t^{\beta_{+}}+c_{2}\,t^{\beta_{-}}\, \,  , \, \,
\beta_{\pm}=\pm\sqrt{\frac{1}{4}-\Omega_{0}^{2}\nu^{2}}\,\,.
\end{equation}

At this stage some comments are in order. In doing so, it turns out to be
meaninful to go back to the original form of the auxiliary equation
(\ref{eqausiliaria}), which now reads
\begin{equation}
\ddot{\sigma}+\Omega_{0}^{2}\,\left(  k_{0}^{2}+\frac{\nu^{2}}{t^{2}}\right)
\sigma-\omega_{0}^{2}\,\frac{1}{4\sigma^{3}}=0\,\,\,.\label{Y19}%
\end{equation}
Next, we consider the equation of motion corresponding to (\ref{malpha}). 
That is (see (\ref{eqtdo}))%
\begin{equation}
\ddot{q}+\frac{\dot{\alpha}}{\alpha}\,\dot{q}+\omega^{2}%
(t)\,q=0\,\,\,,\label{Y24}%
\end{equation}
where $\omega^{2}(t)$ is given by
\begin{equation}
\omega^{2}(t)=\Omega_{0}^{2}\,\left(  k_{0}^{2}+\frac{\nu^{2}}{t^{2}}\right)
+\frac{\ddot{\alpha}}{2\alpha}-\frac{\dot{\alpha}^{2}}{4\alpha^{2}}%
\end{equation}
due to (\ref{defOmega2}) and (\ref{Y3}). By means of the change of variable
$q=y\,/\,\sqrt{\alpha}$ Eq. (\ref{Y24}) is transformed into the equation
\begin{equation}
\ddot{y}+\Omega_{0}^{2}\left(  k_{0}^{2}+\frac{\nu^{2}}{t^{2}}\right)
\,\,y\,=0\,\,\,,\label{nY23}%
\end{equation}
which can be led to a Bessel function. For example, if $\ell=\Omega_{0}k_{0},$
$\Omega_{0}^{2}\nu^{2}=\frac{1}{4}-\rho^{2}$, Equation (\ref{nY23}) is solved
by 
\begin{equation}
y=\sqrt{t}\,Z_{\rho}(\ell\,t)\,\,\,,
\end{equation}
where $Z_{\rho}(t)$ denotes a Bessel function of order $\rho$
defined by the differential equation [\ref{gradshteyn}]
\begin{equation}
\frac{d^{2}}{dt^{2}}Z_{\rho}+\frac{1}{t}\,\frac{d}{dt}Z_{\rho}+\left(
1-\frac{\rho^{2}}{t^{2}}\right)  \,Z_{\rho}=0\,\ \ .\label{nY25}%
\end{equation}
We notice that when the quantity $\frac{1}{4\sigma^{3}}$ is negligible with
respect to the other terms, Eq. (\ref{Y19}) can be approximated by an equation
of the type (\ref{nY23}), namely
\begin{equation}
\ddot{\sigma}+\Omega_{0}^{2}\,\left(  k_{0}^{2}+\frac{\nu^{2}}{t^{2}}\right)
\,\sigma=0\,\,\,.\label{Y21}%
\end{equation}

\section{Discussion}

\setcounter{equation}{0}

We have formulated the uncertainty product $\left(  \Delta_{\alpha}Q\right)
\,\left(  \Delta_{\alpha}P\right)  $ for the generalized (time-dependent)
oscillators (TDO's) with variable mass and frequency, where $\Delta_{\alpha}Q$
and $\Delta_{\alpha}P$ are the variances between extended coherent states
$\left|  \alpha\right\rangle $ of the conjugate position and momentum
operators $Q$ and $P$. The basic starting point of our study is constituted by
the auxiliary equation (\ref{eqausiliaria}) which is associated with the
equation of motion of the TDO under consideration by formula (\ref{sigmasqrt}%
). In the case of the conventional harmonic oscillator, we have simply found
that Eq. (\ref{eq27}) gives rise to two (real) exact different solutions,
(\ref{sigmaset1}) and (\ref{sigmaset2}), whose the first minimizes exactly the
uncertainty expression (\ref{indeterminazioneQP}) ($M=0$, $\dot{\sigma}=0$),
while the second does not minimize exactly this expression, but only
approximately. 

Another case in which the auxiliary equation (\ref{eqausiliaria}) can be
explicitly solved is represented by the particular Kanai-Caldirola oscillator
(\ref{kanaicaldirola}) with $\Omega^{2}(t)=\Omega_{0}^{2}=\omega_{0}%
^{2}-\gamma^{2}/4=const$ and $m(t)=m_{0}\exp\left(  \gamma\,t\right)  $
\thinspace($\gamma=M_{0}$ is the damping). We obtain two possible exact
solutions of the corresponding auxiliary equation accordingly to $\Omega
_{0}^{2}>0$ or $\Omega_{0}^{2}<0$. One solution (for $\Omega_{0}^{2}>0$) is
given by $\sigma=\frac{1}{\sqrt{2 |\Omega_{0}|}}$ 
and, therefore, it resembles
what happens for the usual harmonic oscillator (see (\ref{sigmaset1})), while
the other (for $\Omega_{0}^{2}<0$) is furnished by (\ref{sigmaset3}) and does
yield minimum states only approximately. 

As one expects, the uncertainty relation (\ref{incertezzabogo}) can even be
written as%
\begin{equation}
\sqrt{\left\langle 0\right|  Q^{2}\left|  0\right\rangle \,\left\langle
0\right|  P^{2}\left|  0\right\rangle }=\frac{\hbar}{2}\left[  1+4\sigma
^{2}\left(  \dot{\sigma}-\frac{M}{2}\sigma\right)  ^{2}\right]  ^{\frac{1}{2}%
}\,\,,
\end{equation}
where the expectation values between the vacuum state of $Q^{2}$ and $P^{2}$
are%
\begin{equation}
\left\langle 0\right|  Q^{2}\left|  0\right\rangle =\frac{\hbar}{m}%
\,\sigma^{2}\,\,\,,\quad\left\langle 0\right|  P^{2}\left|  0\right\rangle
=\hbar\,m\,\left|  \xi\right|  ^{2}\,\,\,,
\end{equation}
with $\left|  \xi\right|  ^{2}$ expressed by (\ref{xi2}). Since in the case of
minimum uncertainty $\sigma=c\,m^{1/2}$ (see (\ref{sigmamin})), the
expectation values $\left\langle 0\right|  Q^{2}\left|  0\right\rangle $ and
$\left\langle 0\right|  P^{2}\left|  0\right\rangle $ turn out to be constant,
namely%
\begin{equation}
\left\langle 0\right|  Q^{2}\left|  0\right\rangle =\hbar\,c^{2}\quad
,\quad\left\langle 0\right|  P^{2}\left|  0\right\rangle =
\frac{\hbar}{4\,c^{2}}\,\,\ \ ,
\end{equation}
so that
\begin{equation}
\sqrt{\left\langle 0\right|  Q^{2}\left|  0\right\rangle \,\left\langle
0\right|  P^{2}\left|  0\right\rangle }=\frac{\hbar}{2}%
\end{equation}
corresponds to the minimum uncertainty. In this situation, from (\ref{munu})
we infer that the Bogolubov coefficients $\mu(t)$ and $\nu(t)$ are identical
to those characterizing the usual harmonic oscillator:%
\begin{equation}
\mu(t)=1\,\,\,\,\, ,
\,\,\,\,\,\nu(t)=0\,\,\,\,\,.
\end{equation}
On the other hand, in the light of the criterium (\ref{momegac2}) the
Hamiltonian operator (\ref{HQP}) reads
\begin{equation}
\hat{H}(t)=\frac{m_{0}}{m(t)}\,H(0)
\end{equation}
where $m\omega=m_{0}\omega_{0}=1/(2c^{2})$. We remark that the vacuum state
carries of course zero momentum $\left\langle 0\right|  P\left|
0\right\rangle =0$ but not zero energy, in the sense that the requirement
(\ref{momegac2}) entails%
\begin{equation}
\left\langle 0\right|  \,\hat{H}(t)\,\left|  0\right\rangle =\frac{m_{0}%
}{m(t)}\,\left\langle 0\right|  \,\hat{H}(0)\,\left|  0\right\rangle
=\frac{\hbar\,\omega(t)}{2}\,\,\,\,,\,\,
\end{equation}
which can be interpreted as the zero-point energy of the time-dependent
quantum oscillator minimizing uncertainty states. 

\renewcommand{\thesection}{\Alph{section}} 
\section*{Appendix}
\setcounter{equation}{0} 
\addtocounter{section}{-4}
\setcounter{equation}{0}

In order to solve the nonlinear equation
\begin{equation}
2\alpha\ddot{\alpha}-\dot{\alpha}^{2}-4\omega_{0}^{2}+\,\alpha^{2}\,\left(
\mu^{2}+\frac{\lambda^{2}}{t^{2}}\right)  =0\qquad\label{app1}%
\end{equation}
the series expansion
\begin{equation}
\alpha(t)=\sum_{k=0}^{\infty}a_{k}\,t^{k}\qquad\label{app2}%
\end{equation}
can be introduced. From (\ref{app2}) we thus obtain%
\begin{equation}
\dot{\alpha}^{2}=\sum_{k=1}^{\infty}b_{k}\,t^{k}\quad,\quad\alpha^{2}%
=\sum_{k=0}^{\infty}c_{k}\,t^{k}\quad,\quad\alpha\ddot{\alpha}=\sum
_{k=0}^{\infty}d_{k}\,t^{k}\label{app3}%
\end{equation}
where%
\begin{equation}
b_{k}=\sum_{j=0}^{k}(j+1)\,(k-j+1)\,a_{j+1}\,a_{k-j+1}\quad,\quad c_{k}%
=\sum_{j=0}^{k}a_{j}\,a_{k-j}%
\end{equation}
and%
\begin{equation}
d_{k}=\sum_{j=0}^{k}(2+k-j)\,(1+k-j)\,a_{j}\,a_{k+2-j}\quad.\label{app5}%
\end{equation}
Inserting (\ref{app3}) into (\ref{app1}) and collecting terms of same power of
$t$ yields%
\begin{align*}
&  \lambda^{2}c_{0}=\lambda^{2}c_{1}=0\quad,\\
&  2d_{0}-b_{0}-4\omega_{0}^{2}+\mu^{2}c_{0}+\lambda^{2}\,c_{2}=0\quad,\\
&  2d_{n}-b_{n}+\mu^{2}\,c_{n}+\lambda^{2}\,c_{n+2}=0\quad
\end{align*}
$\forall n\geq1$. In terms of the $a_{i}$'s we have
\begin{align*}
&  \lambda^{2}\,{a_{0}}^{2}= 2\,\lambda^{2}\,{a_{0}}\,{a_{1}}=0\quad,\\
&  4\,{a_{0}}\,{a_{2}}-{a_{1}}^{2}-4\,{\omega_{0}}^{2}+{a_{0}}^{2}\,\mu
^{2}+\lambda^{2}\,({a_{1}}^{2}+2\,{a_{0}}\,{a_{2})}=0\quad,\\
&  12\,{a_{0}}\,{a_{3}}+2\,\mu^{2}\,{a_{0}}\,{a_{1}}+2\,\lambda^{2}\,({a_{1}%
}\,{a_{2}+}\,{a_{0}}\,{a_{3})}=0\quad,\\
&  {6\,{a_{1}}\,{a_{3}}+24\,{a_{0}}\,{a_{4}}+\mu^{2}\,({a_{1}}^{2}%
+2\,\,{a_{0}}\,{a_{2})}+\lambda^{2}(\,{a_{2}}^{2}+2\,{a_{1}}\,{a_{3}%
+2\,\,{a_{0}}\,{a_{4}})}=0}\quad, \\
&  16\,{a_{1}}\,{a_{4}}+4\,{a_{2}}\,{a_{3}}+40\,{a_{0}}\,{a_{5}}+2\,\mu
^{2}\,({a_{0}}\,{a_{3}}+\,{a_{1}}\,{a_{2})}+\\
&  \qquad+\,2\lambda^{2}\,({a_{1}}\,{a_{4}}+{a_{2}}\,{a_{3}+a_{0}}\,{a_{5}%
)}=0\quad,
\end{align*}
\begin{align*}
&  30\,{a_{1}}\,{a_{5}}+12\,{a_{2}}\,{a_{4}}+60\,{a_{0}}\,{a_{6}}+3\,{a_{3}%
}^{2}+\,\mu^{2}\,(2{a_{0}}\,{a_{4}}+2\,\,{a_{1}}\,{a_{3}}+\,{a_{2}}^{2})+\\
&  \qquad+\lambda^{2}\,({a_{3}}^{2}+2\,\,{a_{1}}\,{a_{5}}++2\,\,{a_{2}%
}\,{a_{4}}+2\,\,{a_{0}}\,{a_{6})}=0\quad,\\
&  {48\,{a_{1}}\,{a_{6}}+24\,{a_{2}}\,{a_{5}}+12\,{a_{3}}\,{a_{4}}+84\,{a_{0}%
}\,{a_{7}}+{2\,\mu^{2}\,({a_{0}}\,{a_{5}}}+\,{a_{1}}\,{a_{4}}+\,{a_{2}%
}\,{a_{3})}\,{+}}\\
&  \qquad+{2\,\lambda^{2}(\,{a_{1}}\,{a_{6}}+\,{a_{2}}\,{a_{5}}+\,{a_{3}%
}\,{a_{4}}+{a_{0}}\,{a_{7})}}=0\quad,\\
&  {70\,{a_{1}}\,{a_{7}}+40\,{a_{2}}\,{a_{6}}+22\,{a_{3}}\,{a_{5}}%
+112\,{a_{0}}\,{a_{8}}+8\,{a_{4}}^{2}+{\mu^{2}\,({a_{3}}^{2}+}2\,\,{a_{0}%
}\,{a_{6}}+2\,\,{a_{1}}\,{a_{5}}+}\\
&  \qquad+{2\,\,{a_{2}}\,{a_{4})}+\lambda^{2}\,({a_{4}}^{2}+2\,{a_{1}}%
\,{a_{7}}}+{2\,\,{a_{2}}\,{a_{6}+}}2\,\,{a_{3}}\,{a_{5}+2\,\,{a_{0}}\,{a_{8}%
})}=0\quad,\\
&  {24\,{a_{4}}\,{a_{5}}+96\,{a_{1}}\,{a_{8}}+60\,{a_{2}}\,{a_{7}}+36\,{a_{3}%
}\,{a_{6}}+144\,{a_{0}}\,{a_{9}}+}\mu^{2}(2\,\,{a_{0}}\,{a_{7}}+2\,\,{a_{1}%
}\,{a_{6}}+\\
&  \qquad+2\,\,{a_{2}}\,{a_{5}}+2\,\,{a_{3}}\,{a_{4})}+{\lambda^{2}%
(2\,\,{a_{4}}\,{a_{5}}+2\,\,{a_{1}}\,{a_{8}}+2\,\,{a_{2}}\,{a_{7}+}%
}2\,\,{a_{3}}\,{a_{6}+2{a_{0}}\,{a_{9}})}=0\,\,\,,\\
&  {36\,{a_{4}}\,{a_{6}}+126\,{a_{1}}\,{a_{9}}+84\,{a_{2}}\,{a_{8}}%
+54\,{a_{3}}\,{a_{7}}+180\,{a_{0}}\,{a_{10}}+15\,{a_{5}}^{2}+}\\
&  \qquad+{\mu^{2}\,({a_{4}}^{2}+}2\,\,{a_{0}}\,{a_{8}}+2\,\,{a_{1}}\,{a_{7}%
}++2\,\,{a_{2}}\,{a_{6}}+2\,\,{a_{3}}\,{a_{5})}+\\
&  \qquad+{\lambda^{2}(\,{a_{5}}^{2}}+2\,\,{a_{4}}\,{a_{6}}+2\,\,{a_{1}%
}\,{a_{9}}+2\,\,{a_{2}}\,{a_{8}}+2\,\,{a_{3}}\,{a_{7}+2\,{a_{0}}\,{a_{10}}%
)}=0\quad,\\
&  {52\,{a_{4}}\,{a_{7}}+40\,{a_{5}}\,{a_{6}}+160\,{a_{1}}\,{a_{10}%
}+112\,{a_{2}}\,{a_{9}}+76\,{a_{3}}\,{a_{8}}+220\,{a_{0}}\,{a_{11}}+}\\
&  \qquad+\mu^{2}(2\,{a_{0}}\,{a_{9}}+{2\,\,{a_{4}}\,{a_{5}}}+2\,{a_{1}%
}\,{a_{8}}+2\,\,{a_{2}}\,{a_{7}}+2\,\,{a_{3}}\,{a_{6})+}\\
&  \qquad+{\lambda}^{2}(2\,\,{a_{4}}\,{a_{7}}+2\,\,{a_{5}}\,{a_{6}%
}+2\,\,{a_{1}}\,{a_{10}}+2\,\,{a_{2}}\,{a_{9}}+2\,\,{a_{3}}\,{a_{8}%
+2\,\,{a_{0}}\,{a_{11}})}+\,=0\quad,\\
&  {72\,{a_{4}}\,{a_{8}}+54\,{a_{5}}\,{a_{7}}+198\,{a_{1}}\,{a_{11}%
}+144\,{a_{2}}\,{a_{10}}+102\,{a_{3}}\,{a_{9}}+264\,{a_{0}}\,{a_{12}%
}+24\,{a_{6}}^{2}+}\\
&  \qquad\,+{\nu^{2}\,{a_{6}}^{2}}+2\,\mu^{2}\,({a_{5}}^{2}+{a_{4}}\,{a_{6}%
}+\,{a_{0}}\,{a_{10}}+\,{a_{1}}\,{a_{9}}+\,{a_{2}}\,{a_{8}}+\,{a_{3}}%
\,{a_{7})+}\\
&  \qquad+2{\lambda}^{2}(\,{a_{4}}\,{a_{8}}+\,{a_{5}}\,{a_{7}}+\,{a_{1}%
}\,{a_{11}}+\,{a_{2}}\,{a_{10}}+\,{a_{3}}\,{a_{9}}+\,{a_{0}}\,{a_{12})}=0\quad
\end{align*}
etc. Since $c_{0}=a_{0}^{2}$ then we must require $a_{0}=0$, which also
implies $c_{1}=2a_{0}a_{1}=0$. Moreover, from the third equation we obtain
$a_{1}$ ($\neq0$), that is ${a_{1}}^{2}(\lambda^{2}-1)=4\,{\omega_{0}}^{2}$.
Setting $a_{0}=0$ drastically simplify the whole system in that, due to the
typical bilinear structures involved, the net consequence is that coefficients
$a_{k}$ of the even index must vanish as well. For instance, once we set
$a_{0}=0$ the fourth equation above reads $2\lambda^{2}a_{1}a_{2}=0$. Further
insertion of $a_{2}=0$ into the system results into the sixth equation
$(\lambda^{2}+8)a_{1}a_{4}=0$ and so on. As regards the series coefficients
$a_{k}$ of the odd index, they can be determined recursively from the the
fifth, the seventh etc., equations of the system. These equations now read
\begin{align*}
&  {a_{1}}\,(\mu^{2}\,{a_{1}}+6\,{a_{3}}+2\,\nu^{2}\,{a_{3}})=0\quad,\\
&  30\,{a_{1}}\,{a_{5}}+3\,{a_{3}}^{2}+2\,\mu^{2}\,{a_{1}}\,{a_{3}+\lambda
}^{2}\,({a_{3}}^{2}+2\,\,{a_{1}}\,{a_{5})}=0\quad,\\
&  70\,{a_{1}}\,{a_{7}}+22\,{a_{3}}\,{a_{5}}+\mu^{2}\,({a_{3}}^{2}%
+2\,\,{a_{1}}\,{a_{5})}+2\,{\lambda}^{2}(\,{a_{1}}\,{a_{7}}+\,{a_{3}}%
\,{a_{5})}=0\quad,\\
&  126\,{a_{1}}\,{a_{9}}+54\,{a_{3}}\,{a_{7}}+15\,{a_{5}}^{2}+2\,\mu
^{2}\,({a_{1}}\,{a_{7}}+\,{a_{3}}\,{a_{5})}+{\lambda}^{2}(\,{a_{5}}%
^{2}+2\,\,{a_{1}}\,{a_{9}}+\\
&  \qquad+2\,\,{a_{3}}\,{a_{7})}=0\quad,\\
\lefteqn{54\,a_{5}\,a_{7}+198\,a_{1}\,a_{11}+102\,a_{3}\,a_{9}+\mu
^{2}(2\,\,a_{1}\,a_{9}+2\,\,a_{3}\,a_{7}+\,a_{5}{}^{2})+}\\
&  \qquad+2\lambda^{2}(\,\,a_{5}\,a_{7}+\,a_{1}\,a_{11}+\,a_{3}\,a_{9}%
)=0\quad,\\
\lefteqn{94\,a_{5}\,a_{9}+286\,a_{1}\,a_{13}+166\,a_{3}\,a_{11}+35\,a_{7}%
{}^{2}+2\,\mu^{2}\,(a_{1}\,a_{11}+\,a_{3}\,a_{9}+a_{5}\,a_{7})+}\\
&  \qquad+\lambda^{2}\,(a_{7}{}^{2}+2\,\,a_{5}\,a_{9}+2\,a_{1}\,a_{13}%
+2\,a_{3}\,a_{11})=0\quad,
\end{align*}
etc. It is an easy matter to handle this system iteratively, thus giving all
the odd index coefficients in terms of the $a_{1}$. Precisely, a little bit of
inspection reveals the existence of the recursion relation (\ref{a2k+1a1}),
which can be subsumed even into
\begin{equation}
\frac{a_{2k+1}}{a_{2k-1}}=\,\,\frac{-\mu^{2}\,(2k-1)}{2\,k\,[(4k^{2}%
-1)+\lambda^{2}]}\qquad.
\end{equation}

\section*{References}

\begin{enumerate}
\item \label{ermakov}V.P. Ermakov, Univ. Izv. Kiev \textbf{20}, No. 9, (1880) 1.

\item \label{milne}E.W. Milne, Phys. Rev.\textbf{35}, (1930) 863.

\item \label{pinney}E. Pinney, Proc. Am. Math. Soc. \textbf{1}, (1950) 681.

\item \label{eliezer}C.J. Eliezer and A. Gray, SIAM J. Appl. Math.
\textbf{30}, (1976) 463.

\item \label{espinoza} P.B. Espinoza, \textit{Ermakov-Lewis dynamic invariants 
with some applications}, MS thesis, Universidad de Guanajauto, \textit{math-phys/0002005} 
(2000); and references therein.

\item \label{moya}H. Moya-Cessa and M.F. Guasti, Phys. Lett. \textbf{A311} (2003) 1.

\item \label{goff}S. Goff and D.F. St. Mary, J. Math. Anal. Appl.
\textbf{140}, (1989) 95.

\item \label{lewis}H.R. Lewis,Jr. and W.B. Riesenfeld, J. Math. Phys.
\textbf{10}, (1969) 1458.

\item \label{profilo}G. Profilo and G. Soliani, Phys. Rev. \textbf{A44},
 (1991) 2057.

\item \label{kanai}E. Kanai, Prog. Theor. Phys. \textbf{3}, (1948) 440.

\item \label{caldirola}L. Caldirola, Nuovo Cimento \textbf{B87}, (1983) 241.

\item \label{hartley}J.G. Hartley and J.R. Ray, Phys. Rev. \textbf{D25}, (1982) 382.

\item \label{pedrosa}I.A. Pedrosa, Phys. Rev. \textbf{D36}, (1987) 1279.

\item \label{yuen}H.P. Yuen, Phys. Rev. \textbf{A13}, (1976) 2226.

\item \label{walls}D.F. Walls, Nature, \textbf{306}, (1983) 141.

\item \label{gradshteyn}I.S. Gradshteyn and I.M. Ryzhik, 
\textit{Table of Integrals, Series and Products}, Academic Press, 
New York (1965).
\end{enumerate}
\end{document}